\documentclass[12pt,aps,showkeys,showpacs,preprintnumbers,amsmath,amssymb,nofootinbib]{revtex4}
\usepackage{subfigure}
\usepackage{graphicx}
\usepackage{epstopdf}
\usepackage{color}
\usepackage{amsfonts}
\usepackage{amssymb}
\usepackage{amsmath}
\usepackage{mathtools}
\def \be  {\begin{equation}}
\def \ee  {\end{equation}}
\def \bea {\begin{eqnarray}}
\def \eea {\end{eqnarray}}

\begin{document}
\vspace*{0.5cm}

\title{Particles Multiplicity Based on Rapidity in Landau and Artificial Neural Network(ANN) Models }

\author{D. M. Habashy}
%\email{Emad10@hotmail.com}
\affiliation{Ain Shams University, Faculty of Education, Physics Department, 11771, Roxy, Cairo, Egypt}

\author{Mahmoud Y. El-Bakry}
%\email{Kemo_h3S@hotmail.com}
\affiliation{Ain Shams University, Faculty of Education, Physics Department, 11771, Roxy, Cairo, Egypt}

\author{Abdel Nasser Tawfik}
%\email{atawfik@cern.ch}
\affiliation{Nile University - Egyptian Center for Theoretical Physics (ECTP), Juhayna Square off 26th-July-Corridor, 12588 Giza, Egypt}

\author{R. M. Abdel Rahman}
%\email{R_M_A_15@yahoo.com}
\affiliation{Modern Academy for Engineering and Technology, Basic Sciences Department, 11571, Mokattam, Cairo, Egypt}

\author{Mahmoud Hanafy}
%\email{mahmoud.nasar@fsc.bu.edu.eg}
\affiliation{Physics Department, Faculty of Science, Benha University, 13518, Benha, Egypt}

\date{\today}

\begin{abstract}  

	ANN model is used to estimate the multiplicity per rapidity for charged pions and kaons observed in various high-energy experiments from central Au+Au collisions with energies ranging from 2-200 GeV, and then compared to available experimental data, including RHIC-BRAHMS, and the future facilities at NICA and FAIR. We also used Landau hydrodynamical approach, which has a better describtion for the evolution of hot and dense matter produced in ultra-relativistic heavy-ion collisions. The approach is fitted to both results  estimated from experiment and ANN simulation. We noticed that the Landau model accurately reproduces the entire range of multiplicity per rapidity for all created particles at all energies. Also ANN model can reproduce the multiplicity per rapidity very well for all considered particles.

\end{abstract}

\pacs{24.10.Nz,25.75.-q,05.70.-a}

\keywords{Hydrodynamical models, RHIC, ANN method}

\maketitle
%%%%%%%%%%%%%%%%%%%%%%%%%%%%%%%%%%%%%%%%%%%%%%%%%%%%%
%%%   Section I
%%%%%%%%%%%%%%%%%%%%%%%%%%%%%%%%%%%%%%%%%%%%%%%%%%%%%

\section{Introduction}

The evolution of dense and hot matter produced in ultra-relativistic heavy-ion collisions is an interesting subject \cite{Wong:2008ta,Tawfik:2014eba}. It is important to have a better overview of the dynamics of the matter created in such collision for solving various problems in heavy-ion collisions, such as the research of a heavy quarkonium in quark-gluon plasma(QGP) and the interaction of the produced jet with the formed medium \cite{Niida:2021wut,Ding:2021ajz,Busza:2018rrf}. For these challenges, Landau hydrodynamics model provides a successful picture for the evolution of a dense matter assembly at high temperature and pressure\cite{Du:2021zqz,Wong:2008ta}. Its dynamics can be solved precisely \cite{1965569,Hirano:2002ej,Monnai:2019jkc} during the longitudinal expansion in one-dimension in the initial stage. The longitudinal expansion problem in one-dimension has an approximate solution that can be applied to the fluid in its entirety.
Following that, the three-dimensional behavior can be approximated to get predictions comparable to experimental results \cite{Murray:2004gh,1965569,Blume_2005}. The multiplicity per rapidity is one of the most essential observable in heavy-ion collisions. It can be considered as an important feature for particle production\cite{Bearden:2004yx}.

Many models \citep{HAGEDORN1980136,PhysRevD.91.054025,PhysRevC.85.014908,PhysRevD.27.140,Schnedermann_1993,Braun_Munzinger_1996,Braun_Munzinger_1995,PhysRevC.83.034908,Hirano:2002ej,PhysRevC.66.054904,PhysRevC.56.439,PhysRevC.65.064905,Blume_2005,HAWEEL2003159} have been studied and discussed rapidity distribution, such as Landau hydrodynamical approach, which will be utilized  in this research. Hydrodynamics is employed to describe the evolution of a colliding system, from the initial conditions upto freez-out conditions \cite{Liu:2012jh}.
In 1950, E. Fermi proposed that when high-energy particles collide, numerous new elementary particles are formed, each with a mean free path that is smaller than the total size of the created cloud of hadronic matter \cite{PhysRev.81.683}.

Artificial neural networks (ANNs) are the core components of artificial intelligence, and are frequently involved in machine learning (ML). Firstly, the ML model is built by training the model with a training data set. The performance is assessed using a new set of data, and if necessary, the model parameters are fine-tuned \cite{PANG2021121972,Duarte:2020ngm,Mallick:2021wop}. 
The model is saved and ready to be applied to actual data to solve the problem once the estimations are satisfactory. Machine learning is mostly used to solve classification, regression, and clustering problems \cite{PANG2021121972,brouwer2014investigation}. The challenge we're working on is trained modelling, which means that each pair of input variables has a fixed numerical value as the goal variable \cite{Rojas:1996,article6,article7,DARWISH2015299}. 
In recent years, several successful applications of the Artificial Neural Networks (ANNs) have emerged in nuclear physics and high-energy physics \cite{PANG2019867,PANG2021121972,brouwer2014investigation,Duarte:2020ngm,Mallick:2021wop,Li:2021plq,article3,Apolinario:2021olp,ELBAKRY2003995} , as well as in biology, chemistry, meteorology, and other fields of science. A major goal of nuclear theory is to predict nuclear structure and nuclear reactions using the underlying theory of the strong interactions, Quantum Chromodynamics (QCD), Quark Gluon Plasma and other theories. One of the essential observables that has a substantial impact on the end state particle production is the rapidity distribution of a collision.
Each set of data corresponds to a single heavy-ion collision event \cite{Mallick:2021wop,Li:2021plq,Apolinario:2021olp}. Given that the work is comparable to signal fitting, determining the degree of conformity between the original and obtained signals is crucial. As a result, the performance of the ANNs considered in this study was assessed using a Chi-squared test deformation \cite{Haykin:2008,article5,cryst10040290,Rojas:1996,Stachel:1989pa,NADA201380}. 

The goal of this study is to compute the multiplicity per rapidity for charged pions and kaons formed from Au + Au collisions using the Landau hydrodynamical model and the ANN simulation model at energies ranging from low to high. We compare our results to experimental data from RHIC-BRAHMS \cite{PhysRevLett.90.102301,PhysRevLett.84.5488,PhysRevC.83.034908,Bratkovskaya:2017gxq,BRAHMS:2009acd,BRAHMS:2009wlg,Blume_2005,BRAHMS:2004adc,BRAHMS:2004dwr,E895:2001zms,2009ScChG..52..198F,article4}, as well as future facilities NICA and FAIR. 

This paper is organised as follows. In Sec. (\ref{models}), We briefly introduce Landau hydrodynamical model in Sec. (\ref{sec:LHmodel}) and the ANN model is in Sec. (\ref{sec:ANN}).  The results are shown in In Sec. (\ref{Results and Discussion}). Conclusion is drawn in Sec. (\ref{conc}).
\section{Approaches}
\label{models}

In this section, the particle multiplicity per rapidity is described briefly using Landau approach, which is dependent on the dynamical evolution of dense and hot matter, and the ANN simulation model. According to the present research, these gradients appear to be crucial in particle production. 
  
\subsection{Landau hydrodynamical approach} 
\label{sec:LHmodel}  

A recent analysis of experimental data shows that the Landau hydrodynamical technique produces conclusions that are consistent with experiment \cite{Steinberg:2004wx,Murray:2004gh,Jiang:2013rm, Wong:2008ta}. A quantitative analyses use an approximate form of the Landau rapidity distribution. 

The equation that describes the relationship is based on a Gaussian distribution, which was deduced first from the experimental data. Later, when it became necessary to use and discuss quantitative analysis of hydrodynamic evolution, a series of calculated treatments were carried out, and the equation could be expressed as \cite{Landau:104093,Steinberg:2004vy}
\begin{equation}
\frac{1}{\sigma^2}\frac{d\sigma}{dy}=\frac{dN}{dy}, \label{equ:onee}
\end{equation}

The number distribution of particles is calculated by integrating the single-particle inclusive distribution with respect to the transverse momentum $p_{\bot}$ and then dividing the inelastic cross section \cite{Landau:104093,Steinberg:2004vy}

\begin{equation}
E\frac{d^3N}{d^3p}=\int_{-\infty}^\infty\rho(y_0)E\frac{d^3N_1}{d^3p}(y-y_0)dy_{FB}, \label{equ:twoo}
\end{equation}
 
where 
\begin{equation}
\rho(y_0)=\frac{1}{\sqrt{2}\pi\sigma}\exp(-\frac{y_0^2}{2\sigma^2}),    \label{equ:threee}
\end{equation} 
and
\begin{equation}
\frac{d^3p}{E}=m_{\bot}dm_{\bot}dyd\phi=dp_{\bot}dy,  \label{equ:fourr}
\end{equation}

$\frac{d^3p}{E}$ is the Lorentz invariant momentum space volume element.

The rapidity variable has the advantage of transforming linearly when subjected to a Lorentz transformation as \cite{Landau:104093,Steinberg:2004vy}

\begin{equation}\label{equ:fivee}
E \frac{d^{3} N}{d^{3} p} = E \frac{d^{3} \sigma}{p_{\perp} d p_{\perp} d p_{z} d \phi}=\frac{d^{3} \sigma}{p_{\perp}d p_{\perp} d\left(y-y_{0}\right) d \phi}=\frac{d^{3} \sigma}{m_{\perp} d m_{\perp} d\left(y-y_{0}\right) d \phi},
\end{equation}

\begin{equation}
p_{\bot}dp_{\bot} = m_{\bot}dm_{\bot},  \label{equ:sixx}
\end{equation}
and
\begin{equation}
d(y-y_0)=\frac{dp_{\bot}}{E},  \label{equ:sevenn}
\end{equation}

Eq. (\ref{equ:fivee}) can be rewritten as a function of rapidity distribution as:
\begin{equation}
E\frac{d^3\sigma}{d^3p}=\frac{d^3N}{m_{\bot}dm_{\bot}d(y-y_0)d\phi}=\frac{d^2N}{2\pi m_{\bot}dm_{\bot}d(y-y_0)}=\frac{d^2N}{2 \pi p_{\bot}dp_{\bot}d(y-y_0)}, \label{equ:eightt}
\end{equation}

The phase-space distribution function is given by

\begin{equation}
E\frac{d^3\sigma}{d^3p} \equiv f(E,p_z), \label{equ:ninee}
\end{equation}

A link between beam energy and the number of created charged particles per pair of participants is discovered by Landau \cite{Landau:104093}. 
The rapidity distribution of charged particles created is written as \cite{Landau:104093,Steinberg:2004vy}

\begin{equation}
\frac{dN}{dy}\simeq\exp(\sqrt{L^{2} - y^{2}}),
\label{equ:tenn}
\end{equation}

Where $L$ is the logarithm of Lorentz contract factor $\gamma$ \cite{Landau:104093,Steinberg:2004vy}, and it consider as the measure of the thickness of the lorentz contracted disks of the colliding hadronic matter \cite{Landau:104093,Steinberg:2004vy}

\begin{equation}
L=\ln(\gamma)= \ln(\frac{\sqrt{s_{NN}}}{2m_{p}})= \sigma_{y}^{2},
\label{equ:elevenn}
\end{equation}

where $\sqrt{s_{NN}}$ is the center of mass energy, $\sigma_{y}$ is the width of the distribution, and $m_{p}$ is the mass of proton \cite{Cleymans:2007jj}.

Tha Landau Gaussian distribution of the rapidity distribution can be expressed as: \cite{Landau:104093,Stachel:1989pa,Netrakanti:2005iy,Wong:2008ta,Jiang:2013rm}

\begin{equation}
\frac{dN}{dy}= \frac{N}{\sqrt{2\pi L}}\exp(-\frac{y^{2}}{2L}).
\label{equ:twelvee}
\end{equation}

where $N$ is the normalization constant.

However this equation is an approximate representation of Eq. (\ref{equ:onee}) in the region of $|y|<< L$. In other rapidity regions, the distributions are totally different \cite{Netrakanti:2005iy,Jiang:2013rm}.

Tab.(\ref{tab:1}) includes the values of the fitting parameters $N$ and $L$ according to the different used center of mass collision energies for all particles.

\subsection{Artificial Neural Network (ANN) method}
\label{sec:ANN}

Artificial neural networks (ANNs) are one of the most successful programming paradigms in the last two decades. It is widely used in a large variety of applications in various areas because of its great capability and excellent learning function\cite{Akkoyun:2019kve,Anil:2020lch,Duarte:2020ngm,Li:2021plq}. An artificial neural network may theoretically approximate arbitrary continuous mapping with arbitrary precision. By learning \cite{article5,article6}, an artificial neural network can learn desired job \cite{Pandey_2016,article2}.

\begin{figure}[htb]
\includegraphics[width=0.4\linewidth]{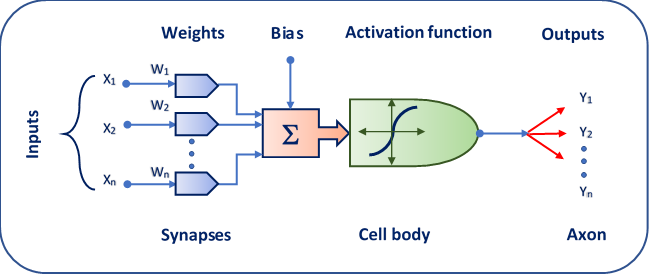}
\caption{The ANN data processing for a neuron iput signal to be processeed and converted to the desired output signal. }
\label{fig:one}
\end{figure}

It's a computer model based on the structure and behaviour of biological neural networks in the brain \cite{Rosenblatt:1958}. While mathematical algorithms are well suited for linear programming, arithmetic and logic calculations, ANNs are more effective to solve problems related to pattern recognition, matching, clustering and classification \cite{Haykin:2008,Gurney:1997}. ANN consists of a very large number of nerve cells (in humans about ten billions nerve), called artificial neurons,as a nonlinear processing units, which are linked to one another in a complex network via synaptic weights\cite{Rojas:1996}. The intelligent behaviour is the outcome of a large amount of interaction among interconnected parts. The input of a neuron is composed of the output signals of the neurons connected to it\cite{article7}. When the contribution of these inputs exceeds a certain threshold, the neuron generates a bioelectric signal through a suitable transfer function, which propagates through the synaptic weights to other neurons \cite{article3} as shown in fig. (\ref{fig:one}).

	For each an artificial neuron, an input $X_n$ is weighted by a factor $W_n$ and added to bias $b$ as 
	 \begin{equation}
	Y_n= \sum_{i}^{n}[X_n W_n + b]. \label{yyyyyy}
	\end{equation}
	To get the output $Y_n$, this problem is pretty hard to solve using the conventional programming approach \cite{Rojas:1996}. There are two types of models: supervised and unsupervised. A supervised model requires a “teacher” or desired output to learn a task, while an unsupervised model does not require a “teacher,” it does learn a task based on the task's objective functions. The development of the first ANN was based on a very simple model of neural connections. The machine $Mark I Perceptron$, which introduced by the neurobiologist Rosenblatt. He assumes that the artificial connections between neurons could change through a supervised learning process \cite{Rosenblatt:1958} that reduces the misfit between actual and expected output. The expected output comes from a set of training data. The misfit between the actual and expected responses of the network represents the necessary information for improving the learning performance \cite{article6}.

	ANN can learn a task by adjusting weights. Automatically learning from data\cite{article6}. Neural network is used to excute different stages of processing systems based on learning algorithms by controlling their weights and biases \cite{article7}. The models excuted by neural networks and cannot be explained in human symbolic language. The result must be accepted as a black box. Expressly, a neural network is able to generate a valid result or being acceptable, but it is not possible to explain how and why this result has been proceed.

	Furthermore, ANNs can achieve solutions that are either algorithmic, computationally costly, or do not exist.
As a result, ANNs are seen as a more powerful modelling technique for advanced complicated nonlinear input-output situations. Mean Square Error (MSE) is the most popular performance procedure for training ANNs, and it has been widely employed in a range of issues such as pattern recognition and machine learning. MSE is described as:

\begin{equation}
\text{MSE} = \sum_{i=1}^{n} [\frac{(y_s-y_o)^2}{n}].   \label{equ:thirteenn}
\end{equation}
where $n$ is the number of datasets used for training the network, $y_s$ is the mean of the simulated value, and $y_o$ is the corresponding experimental value.

	The datasets used for the training, testing and validation purpose of the model are randomized \cite{Haykin:2008,article5,article6}. 
A hyperbolic tangent sigmoid function (tansig) and logarithmic sigmoid function (logsig) are the most commonly utilized activation function in hidden layers , their mathematical formula appears in Eqs. (\ref{equ:fourteenn}) and  (\ref{equ:fifteenn}), whereas a pure linear function (purelin) is utilized in the output layer \cite{Pandey_2016}. 

A = purelin(N,FP) takes N and optional parameters for the function,

\begin{equation}
(logsig)f(x)= \frac{e^x-e^{-x}}{e^x+e^{-x}}.	\label{equ:fourteenn}	
\end{equation}

\begin{equation}
(tansig)f(x) = \frac{1}{1+e^{-x}}.		\label{equ:fifteenn}
\end{equation}

The Pureline function is a neural transfer function that calculate a layer’s output from its net input and its formula expressed as 

\begin{equation}
A = purelin(N, FP).							\label{equ:sixteenn}	
\end{equation}

where $A$ is the output, $N$ is S-by-Q matrix of net input vectors, and $FP$ is struct of function parameters.
	When training the ANN model, different sets of internal network parameters were utilised to define the number of hidden layers, the number of neurons in the hidden layer, and the transfer function.
The number of neurons in the hidden layer is an important consideration when choosing a neural network architecture, and their number varies from case to case. 

In the hidden layers, using sigmoid transfer functions, sometimes known as "squashing" functions, is particularly efficient. They reduce a large input range to a small output range. Sigmoid functions are distinguished by their slope, which must approach zero as the input size increases. When steepest descent is used to train a multilayer network with sigmoid functions, this results in tiny changes in the weights and biases, even though they are far from their optimal values.

\begin{figure}[htb]
\includegraphics[width=0.4\linewidth]{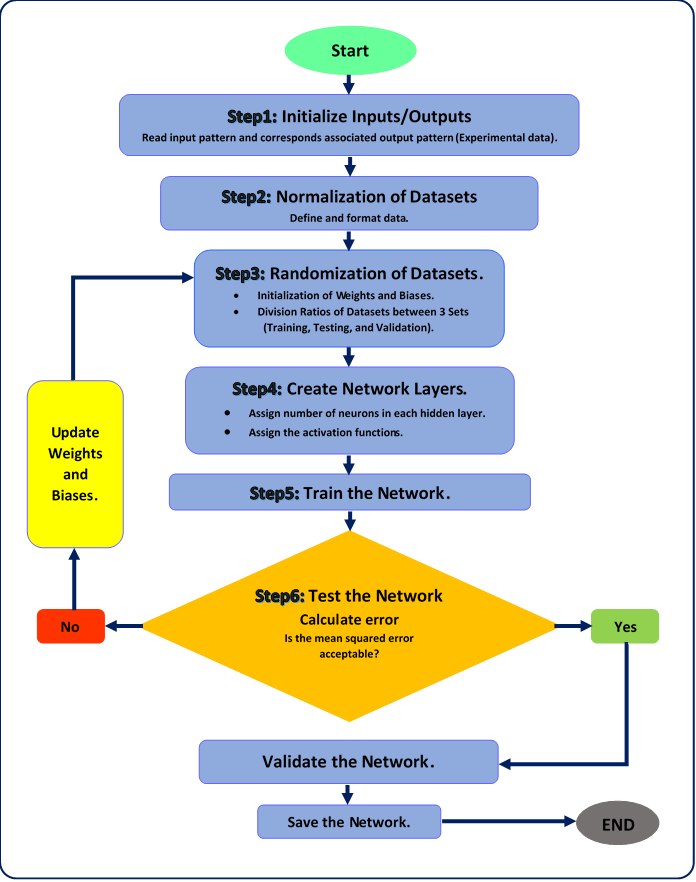}
\caption{A typical flow-chart for ANN method work mapping. }
\label{fig:two}
\end{figure}

	The present work uses a Back Propagation ANN method and $Rprop$ training data to estimate the rapidity distribution $dN/dy$ for particles $\pi^{-}$, $\pi^{+}$, $K^{-}$, and $K^{+}$ produced from (Au + Au) nucleus collision at energies ranging from $2 - 200$ GeV from various experimental high energy physics data of differnt experiments, including RHIC-BRAHMS, and the future facilities NICA and FAIR. A schematic flow-chart to describe the methodology of ANN method is shown in fig.(\ref{fig:two}).

	The used algorithm is $Resilient back-propagation algorithm$ (Rprop). According to this algorithm, we use the trainrp function for training process. Trainrp is a network training function that depends upon make updates for weights and biases. The Rprop algorithm intoduced by both of Riedmiller and Braun as first-order learning methods for neural networks.It considered one of the best performing ANN learning methods.

	The purpose of the $Resilient backpropagation$ (Rprop) training algorithm is to eliminate these harmful effects of the magnitudes of the partial derivatives of arbitrary error measure that is differentiable with respect to the weights. The direction of the weight update is determined solely by the sign of the derivative; the size of the derivative has no bearing on the weight update. A separate update value determines the size of the weight change. 
	
Simulation is done by using Matlab programming language interface as shown in fig. (\ref{fig:three}) for all considered particles. The inputs parameters are rapidity sets, the center of mass energy and the mass number of the gold nucleus (Au), $(Z)$. For a better rapidity distribution, $dN/dy$, the used hidden layers are four, each one includes 100 neurons and the first starting applying epochs (refers to one cycle through the full training dataset)is 1000. Usually, training a neural network takes more than a few epochs. The suitable activation function used for hidden layers is $logsig$ while $purelin$ function used for the output layer.

\section{Results and Discussion}
\label{Results and Discussion}  
	In the present research, we calculate the rapidity distribution, $dN/dy$, produced from various high energy experimental results in central (Au + Au) at energies spanning from $2 - 200$ GeV, including RHIC-BRAHMS, for particles $\pi^{-}$, $\pi^{+}$, $K^{-}$, and $K^{+}$ \cite{PhysRevLett.90.102301,PhysRevLett.84.5488,PhysRevC.83.034908,Bratkovskaya:2017gxq,BRAHMS:2009acd,BRAHMS:2009wlg,Blume_2005,BRAHMS:2004adc,BRAHMS:2004dwr,E895:2001zms,2009ScChG..52..198F,article4}. The obtained results are confronted to those estimated from ANN simulation model \cite{Rojas:1996,Gurney:1997}. 
	
The ANN model is perfectly trainned, based on the avaliable experimental data, using the input parameters which are shown in Tab. (\ref{tab:2}).

\begin{table}[htb]
\caption {ANN paramters which are used for estimating the multiplicity per rapidity of all considered particles.}  
\centering
\begin{tabular}{|c|c c|c c|c c|c c|} 
\hline 
\hline 
$ANN$&&&&Particles&&&&\\
\cline{2-9}
  parameters & &$\pi^{-}$  & & $\pi^{+}$ & & $K^{-}$ & & $K^{+}$ \\ 
\hline 
\hline 
 $Inputs$ & & $y$ , $\sqrt{{S}_{NN}}$, $Z$  & & $y$ , $\sqrt{{S}_{NN}}$,  $Z$ & & $y$ , $\sqrt{{S}_{NN}}$,  $Z$ & & $y$ , $\sqrt{{S}_{NN}}$,  $Z$   \\
\hline 
$\sqrt{{S}_{NN}}$  & & $2,4,6,8,8.8,$ & & $2,4,6,8,$& &$10.7,62.4,200$ & &$8.8,12.2,17.3,$\\ 
(GeV)&&$12.2,17.3,62.4,200$ & & $10.7,62.4,200$ & & && $62.4,200$\\
\hline
$Z$ &&$Z_{Au}$=197 &&$Z_{Au}$=197 &&$Z_{Au}$=197 &&$Z_{Au}$=197 \\
\hline
$Outputs$ & & $dN/dy$  & & $dN/dy$ & & $dN/dy$ & & $dN/dy$  \\
\hline 
 Hidden layers & & 4 & & 4 & & 4 & & 4   \\
\hline 
 Neurons & & 100,100,100,100  & & 100,100,100,100 & & 100,100,100,100 & & 100,100,100,100  \\
\hline 
 Epochs & & 1000  & & 1000 & & 227 & & 456   \\
\hline 
 Training algorithms & & Rprop  & & Rprop & & Rprop & & Rprop   \\
\hline 
 Training functions & & trainrp  & & trainrp & & trainrp & & trainrp   \\
\hline 
Transfer functions & & $logsig$  & & $logsig$ & & $logsig$ & & $logsig$   \\
\hline 
Performances & &  $0.00136$  & &  $0.00233$  & &  $9.19 \times 10^{-6}$   & &  $9.75 \times 10^{-6}$      \\
\hline 
Output functions & & $ Purelin $  & & $ Purelin $ & & $ Purelin $ & & $ Purelin $   \\
\hline
\hline
 \end{tabular}
 \label{tab:2}
\end {table}

\begin{figure}[htb]
\includegraphics[width=0.35\linewidth]{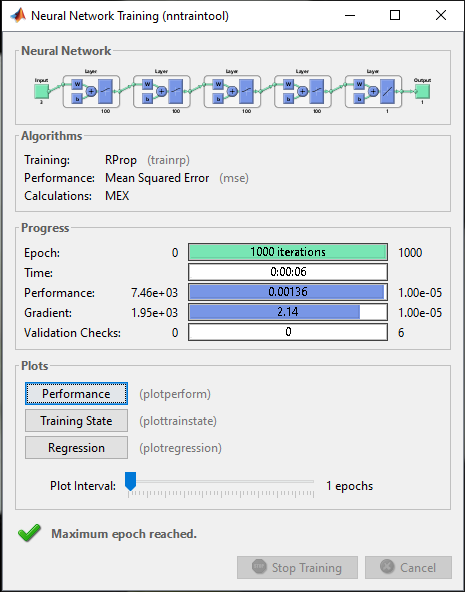} 
\includegraphics[width=0.35\linewidth]{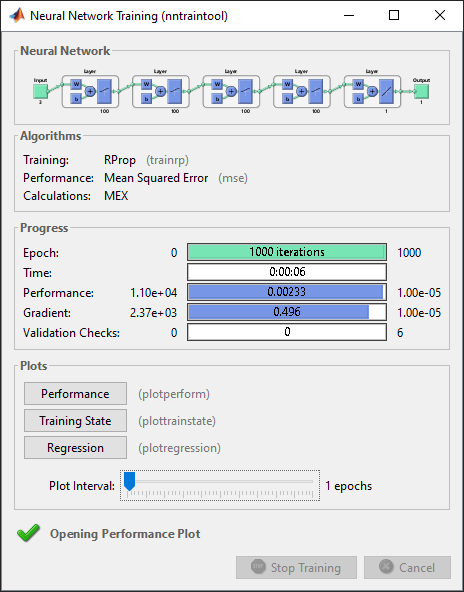}
\includegraphics[width=0.35\linewidth]{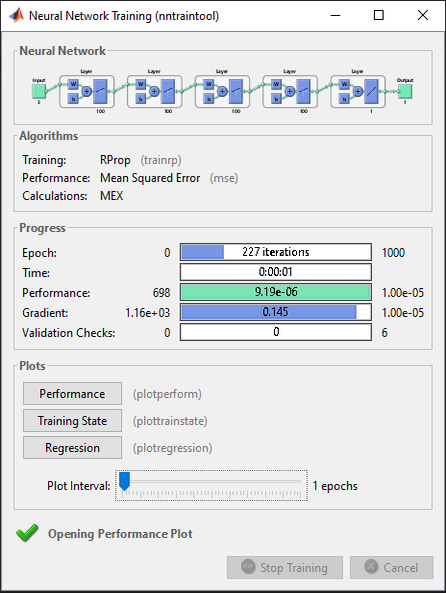}
\includegraphics[width=0.35\linewidth]{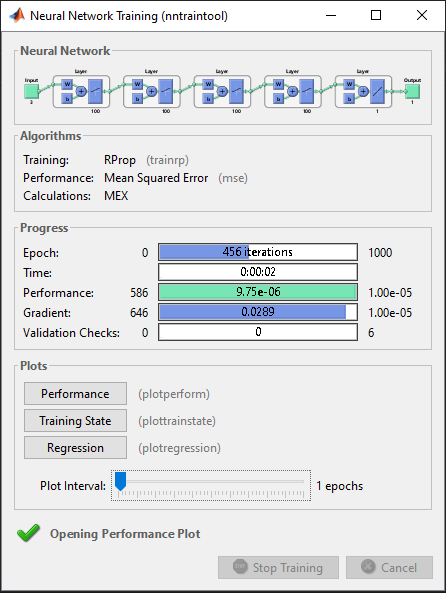}
\caption{Matlab neural network training tool used for estimating the rapidity distribution, $dN/dy$, for particles $\pi^{-}$, $\pi^{+}$, $K^{-}$, and $K^{+}$.}
\label{fig:three}
\end{figure}

\begin{figure}[htb]
\includegraphics[width=0.35\linewidth]{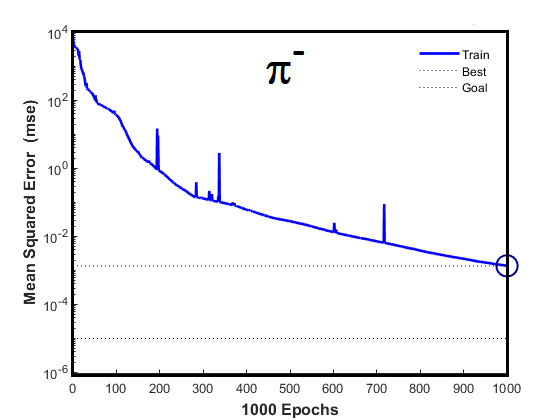} 
\includegraphics[width=0.35\linewidth]{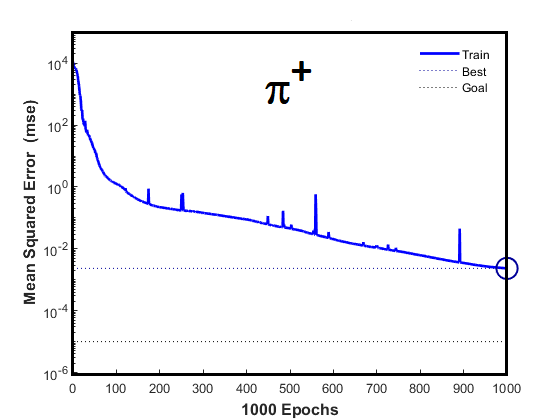} 
\includegraphics[width=0.35\linewidth]{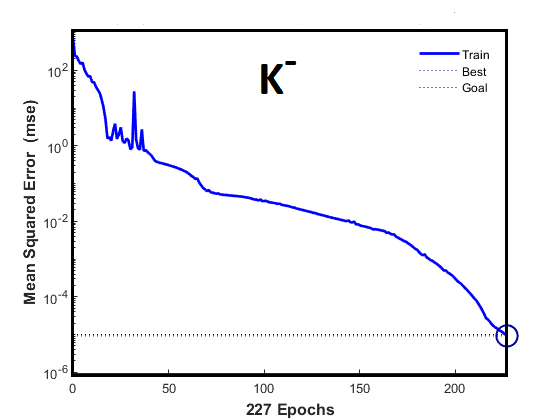} 
\includegraphics[width=0.35\linewidth]{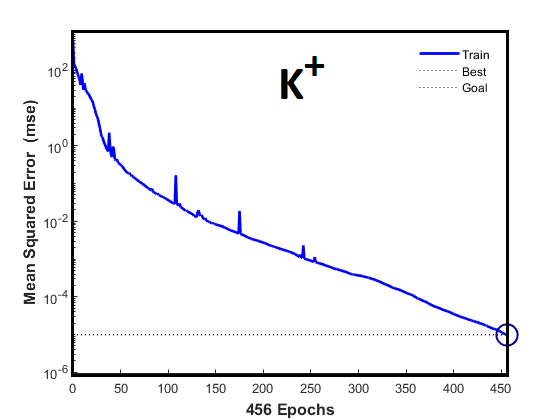} 
\caption{The performance of the used neural network for partciles  (a) $\pi^{-}$, (b) $\pi^{+}$, (c) $K^{-}$, and (d) $K^{+}$.}
\label{fig:four}
\end{figure}

 The ANN target to get the best MSE value about ($10^{-5}$) according to eq.(\ref{equ:thirteenn}). The obtained training results for particles $\pi^{-}$, $\pi^{+}$, $K^{-}$, and $K^{+}$ are shown in figs. (\ref{fig:three}) and (\ref{fig:four}). The network's mean squared error value reduced from a huge value to a lower value, as can be shown from the figure. Furthermore, the network was evolving. after the network has 
learned the training set, training was finalised. The optimal ANN simulation model training, for particles $\pi^{-}$, $\pi^{+}$, $K^{-}$, and $K^{+}$, is choosen as a result of MSE value of $0.00136$, $0.00233$, $9.19 \times 10^{-6}$, and $9.75 \times 10^{-6}$, respictively. The MSE value for pions particles is slightly high in comparison with that of kaons despite of considering the same trainig parameters. This encourage for further investigation.	

The rapidity distribution equation $dN/dy$ can be estimated from ANN as:

\begin{equation}\label{equ:seventeenn}
\begin{array}{l}
d N / d y=\text { purelin }[\text { net. } L W\{5,4\} \text { logsig }(\text { net. } L W\{4,3\} \text { logsig }(\text { net. } L W\{3,2\} \\
\text { logsig }(\text { net. } L W\{2,1\} \text { logsig }(\text { net.IW }\{1,1\} R+\text { net.b }\{1\}) \\
\quad+\text { net.b }\{2\})+\text { net.b }\{3\})+\text { net.b }\{4\})+\text { net.b }\{5\}]
\end{array}
\end{equation}

Where $R$ contains the input parameters,( rapidity, center of mass energy, and Au mass number), which is used to calculate $d N/d y$. $\text {IW}$ and $\text{LW}$ are the linked weights where $\text { net. } I W\{1,1\}$, $\text { net. } L W\{2,1\}$, $\text { net. } L W\{3,2\}$, $\text { net. } L W\{4,3\}$, and $\text { net. } L W\{5,4\}$ reprsent the linked weights between the input layer and first hidden layer, first and second hidden layer, the second and third hidden layer, the third and fourth hidden layer, and the fourth and output layer, respectively.  
The parameter $b$ stands for the bias where $\text { net.b }\{1\}$, $\text { net.b }\{2\}$, $\text { net.b }\{3\}$, $\text { net.b }\{4\}$, and $\text { net.b }\{5\}$ represents the bias of the, first, second, third, and fourth, hidden layer, and the output layer, respectively. 

After that, the rapidity distribution, $dN/dy$, obtained from Landau hydodynamical analysis, according to eq. (\ref{equ:twelvee}), is fitted to the experimental data and compred to those obtained from the ANN simulation model.

The present work aims to study and estimate the rapidity distribution, $dN/dy$, in the frame work of Landua hydrodynamical model based on Eq. (\ref{equ:twelvee}) that ensures a Gaussian distribution for all used particles. The rapidity distribution's dependence $dN/dy$ as function of rapidity $y$ is fitted to a Gaussian distribution. The best fitting parameters are shown in Tab. (\ref{tab:1}).

\begin{table}[htbp]
\caption {The Landau hydrodynamical approach fit parameters for multiplicity per rapidity $dN/dy$ for partciles $\pi^{-}$, $\pi^{+}$, $K^{-}$,and $K^{+}$ at the considered energies.}  
\centering
\begin{tabular}{|c|c|c|c|} 
\hline 
 $particle$ & $\sqrt{{S}_{NN}}$ GeV & $N$ & $L$  \\ 
\hline 
 & $2$   & $32.91 \pm 0.75 $ & $0.34 \pm 0.07$  \\
\cline{2-4} 
 & $4$   & $71.31 \pm 3.1 $ & $0.52 \pm 0.08$  \\
\cline{2-4} 
 & $6$   & $122.07 \pm 3.06 $ & $0.93 \pm 0.08$  \\
\cline{2-4} 
 & $8$   & $143.19 \pm 5.76 $ & $0.948 \pm 0.07$  \\
\cline{2-4} 
 $\pi^{-}$ & $8.8$   & $316.78 \pm 11.17 $ & $1.21 \pm 0.08$  \\
\cline{2-4} 
  & $10.7$   & $146.66 \pm 10.1 $ & $0.75 \pm 0.05$  \\
 \cline{2-4} 
 & $12.2$   & $478.95 $ & $1.591 \pm 0.06$  \\
 \cline{2-4} 
 & $62.4$   & $1025.17 \pm 10.76 $ & $3.289 \pm 0.05$  \\
 \cline{2-4} 
 & $200$   & $1809.23 \pm 12.97 $ & $ 5.4785 \pm 0.07 $  \\ 
 \cline{2-4}
\hline
 & $2$   & $18.091 \pm 0.43 $ & $ 0.328 \pm 0.05 $ \\
 \cline{2-4}
 &   $4$ & $44.05 \pm 2.671 $ & $0.402 \pm 0.06 $  \\
\cline{2-4}
 &   $6$ & $95.55$ & $0.93 \pm 0.06 $  \\
 \cline{2-4}
 $\pi^{+}$ & $8$ & $110.37 \pm 7.01 $ & $1.11 \pm 0.054 $  \\
 \cline{2-4}
 &   $10.7$ & $130 \pm 6.17 $ & $0.84 \pm 0.03 $  \\
 \cline{2-4}
 &   $62.4$ & $1044.02$ & $3.3588 \pm 0.06 $  \\
 \cline{2-4}
 &   $200$ & $1694.12 \pm 12.65 $ & $5.1688 \pm 0.03 $  \\ 
 \cline{2-4}
\hline
&   $10.7$ & $3.92 \pm 0.581 $ & $0.59 \pm 0.06 $  \\
 \cline{2-4}
$K^{-}$ & $62.4$ & $119.223 \pm 0.06 $ & $2.4779 \pm 0.04 $  \\
 \cline{2-4}
 &   $200$ & $226.753 \pm 0.46 $ & $4.2672 \pm 0.03 $  \\ 
 \cline{2-4} 
 \hline
 & $8.8$   & $61.39 $ & $1.211 \pm 0.05$  \\
\cline{2-4} 
 & $10.7$   & $23.44 \pm 0.08 $ & $0.85 \pm 0.05$  \\
\cline{2-4} 
 $K^{+}$ & $12.2$   & $83.241 \pm 0.5 $ & $1.951 \pm 0.05$  \\
 \cline{2-4} 
 & $17.3$   & $112.38 \pm 0.05 $ & $2.06 \pm 0.05$  \\
 \cline{2-4} 
 & $62.4$   & $163.777 \pm 0.06 $ & $3.5056\pm 0.07$  \\
 \cline{2-4} 
 & $200$   & $280.372 \pm 0.03 $ & $5.6778 \pm 0.05$  \\ 
 \cline{2-4}
 \hline
 \end{tabular}
 \label{tab:1}
\end {table}

\begin{figure}[htb]
\includegraphics[width=0.35\linewidth]{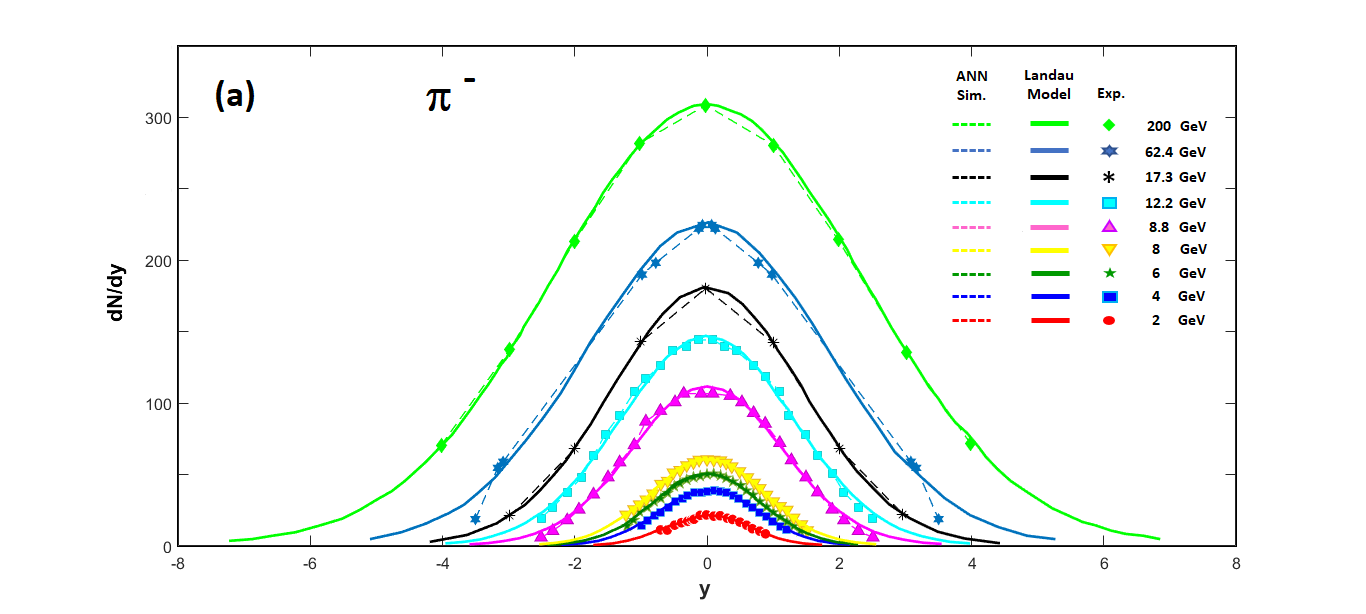}
\includegraphics[width=0.35\linewidth]{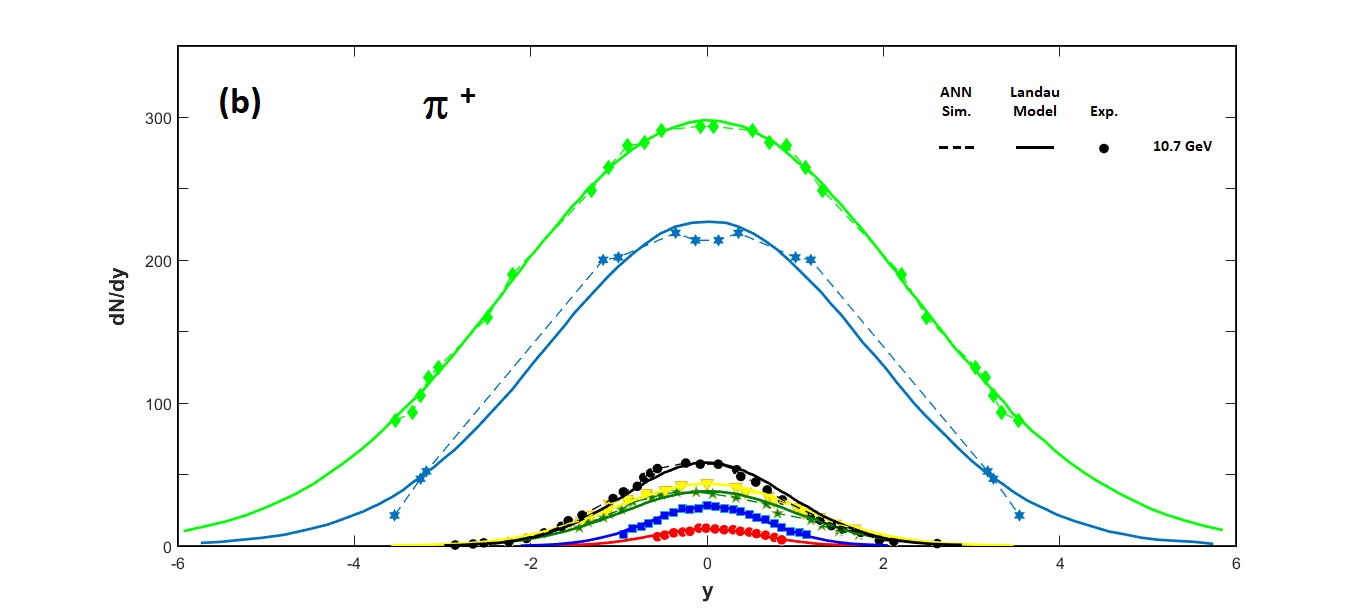}
\includegraphics[width=0.35\linewidth]{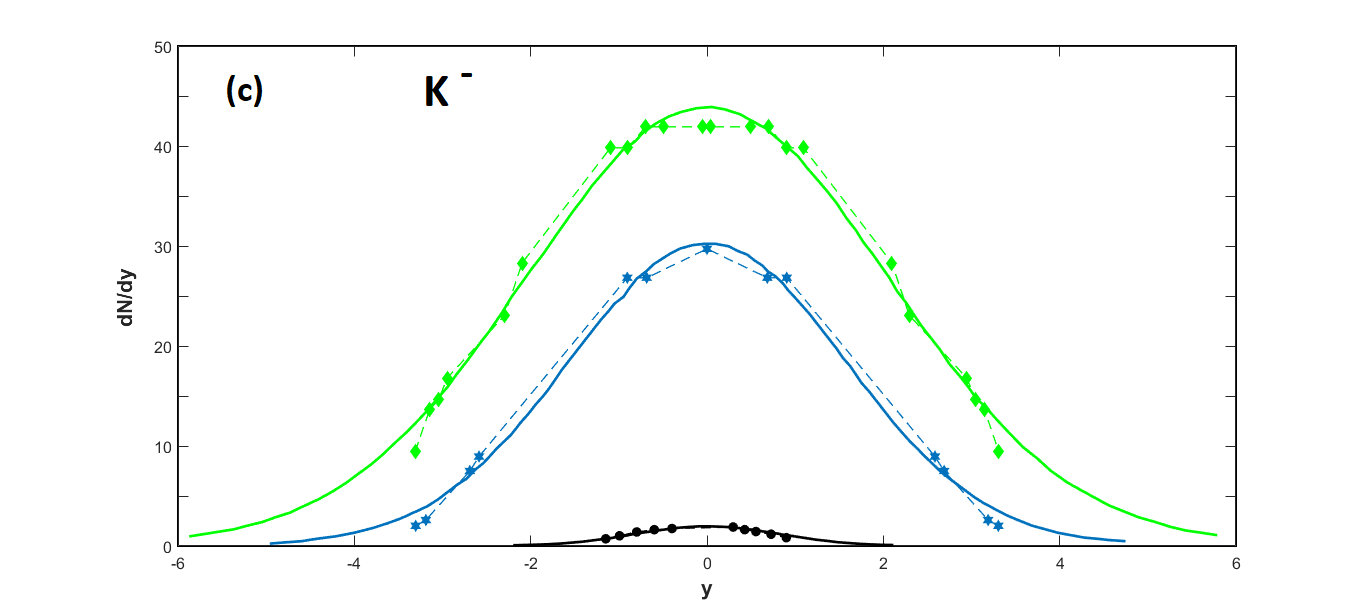}
\includegraphics[width=0.35\linewidth]{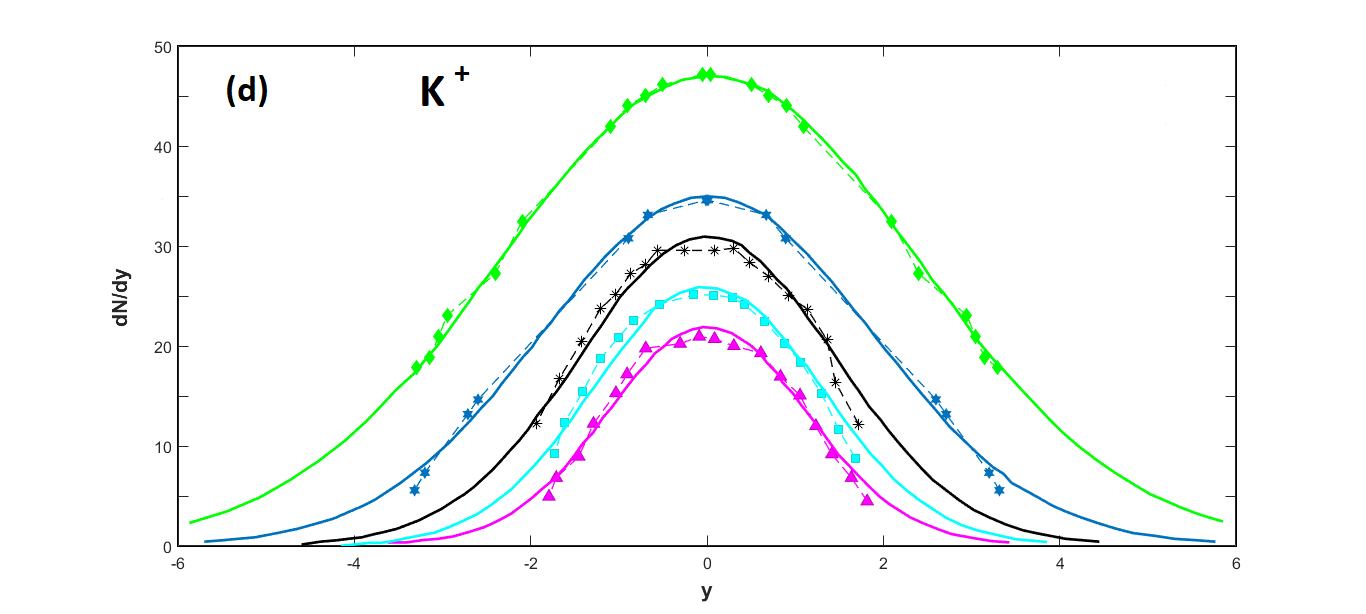}
\caption{The rapidity distribution $dN/dy$ is ploted as a function of the rapidity, $y$, for particles (a) $\pi^{-}$, (b) $\pi^{+}$, (c) $K^{-}$, and (d) $K^{+}$. The experimental results (symbols) from Au+Au central collisions, at $2$, $4$, $6$, $8$, $8.8$, $12.2$, $17.3$, $62.4$, and $200$ GeV \cite{PhysRevLett.90.102301,PhysRevLett.84.5488,PhysRevC.83.034908,Bratkovskaya:2017gxq,BRAHMS:2009acd,BRAHMS:2009wlg,Blume_2005,BRAHMS:2004adc,BRAHMS:2004dwr,E895:2001zms,2009ScChG..52..198F,article4}, are confronted to numerical simulation by using ANN method (solid lines) according to Eq. (\ref{equ:seventeenn}), and also fitted to the Landau hydrodynamical approach (dashed lines), Eq. (\ref{equ:twelvee}).}
\label{fig:threee}
\end{figure}

Fig. (\ref{fig:threee}) shows the rapidity distribution, $dN/dy$, vs rapidity, $y$, for particles $\pi^{-}$, $\pi^{+}$, $K^{-}$, and $K^{+}$ measured in (Au+Au) central nuclear collisions, including RHIC-BRAHMS and the future NICA and FAIR experiments results, at energies $2$, $4$, $6$, $8$, $8.8$, $12.2$, $17.3$, $62.4$, and $200$ GeV \cite{PhysRevLett.90.102301,Murray:2004gh} at a wide rapidity range, $-8 < y < 8$. Experimental data are represented as symbols, while the Landau hydrodynamical model results are represented as solid lines, and the result from ANN simulation method, using Matlab neural network training tool(nntraintool), are depicted as dashed lines. 

We observed that the estimated rapidity distribution $dN/dy$, for particles $\pi^{-}$, $\pi^{+}$, $k^{-}$, and $k^{+}$, exhibit a Gaussian distribution shape. The estimated rapidity distribution, $dN/dy$, around the mid-rapidity ($y=0$), is increased while going from low center of mass energy, $\sqrt{S}=2$ GeV, upto high center of mass energy, $\sqrt{S}=200$ GeV. Accordingly, as the center of mass energy increased the curves shifted upward \cite{Blume_2005}. A general observation is Landau model, which presented in Eq. (\ref{equ:twelvee}), can reproduce the experimental data at the considered range of rapidity, successfully. Also, the ANN simulation method results seem to coincide with the used experimental data \cite{PhysRevLett.90.102301,PhysRevLett.84.5488,PhysRevC.83.034908,Bratkovskaya:2017gxq,BRAHMS:2009acd,BRAHMS:2009wlg,Blume_2005,BRAHMS:2004adc,BRAHMS:2004dwr,E895:2001zms,2009ScChG..52..198F,article4}. In case of $K^{-}$, there is a lake of the experimental data. Only results of the rapidity distribution $dN/dy$ for , $K^{-}$, are avaliable at energies $10.7$, $62.4$, and $200$ GeV measured in (Au+Au) central collisions at various experimental high energy physics \cite{PhysRevLett.90.102301,Murray:2004gh}, including RHIC-BRAHMS and the future experimental facilities, NICA and FAIR. These are general observations that can be used to draw conclusions.

The multiplicity per rapidity, $dN/dy$, for partciles $\pi^{-}$, $\pi^{+}$, $K^{-}$,and $K^{+}$, measured from (Au+Au) central nuclear collisions from various experimental data, including RHIC-BRAHMS and the future experimental facilities FAIR and NICA, at energies $2$, $4$, $6$, $8$, $8.8$, $12.2$, $17.3$, $62.4$, and $200$ \cite{PhysRevLett.90.102301,PhysRevLett.84.5488,PhysRevC.83.034908,Bratkovskaya:2017gxq,BRAHMS:2009acd,BRAHMS:2009wlg,Blume_2005,BRAHMS:2004adc,BRAHMS:2004dwr,E895:2001zms,2009ScChG..52..198F,article4}, is compared to those obtained from the ANN simulation model. The rapidity range considered lies in ($ -8 < y < 8$). Also, the outcomes of the Landau hydrodynamical approach is fitted to the used experimental data and compared to the results of ANN simulation model. The ANN method can reproduce the rapidity distribution, $dN/dy$, at both low and high energies. The low energy region shall be covered by the future NICA and FAIR facilities.

\section{Conclusions}
\label{conc}

We have estimated and calculated the multiplicity per rapidity, $dN/dy$, for partciles $\pi^{-}$, $\pi^{+}$, $K^{-}$,and $K^{+}$, measured from (Au+Au) central nuclear collisions from various experimental data, including RHIC-BRAHMS and the future experimental facilities FAIR and NICA using two different models, namely the ANN simulation model; the most successful single programming paradigms ever invented in the last two decades and the Landau hydodynamical model. 
The Landau hydrodynamical model is fitted to the considered experimental data and compared to the results of the ANN simulation model. The comparison between the simulation and the measurements shows an excellent agreement. Also, the rapidity distribution and the considered rapidity range is successfully recalculated using the Landau hydrodynamical model. Mathematical expression described experimental data was obtained using the ANN. 

%%%%%%%%%%%%%%%%%%%%%%%%%%%%%%%%%%%%%%%%%%%%%%%%%%%%%%%%%%%%%%%%%%%%%%
%%%   References
%%%%%%%%%%%%%%%%%%%%%%%%%%%%%%%%%%%%%%%%%%%%%%%%%%%%%%%%%%%%%%%%%%%%%%

\bibliographystyle{aip}
\bibliography{Reham_2021}

\end{document}